# Multimorphism and gap opening of charge-density-wave phases in monolayer VTe$_2$

Meizhuang Liu[1,⊥], Changwei Wu[1,⊥], Zizhao Liu[1], Dao-Xin Yao[1,*] and Dingyong Zhong[1,*]

[1]School of Physics and State Key Laboratory of Optoelectronic Materials and Technologies, Sun Yat-sen University, 510275 Guangzhou, China

[⊥]These authors contributed to the work equally

*e-mail: yaodaox@mail.sysu.edu.cn; dyzhong@mail.sysu.edu.cn

**Abstract**

Vanadium dichalcogenides have attracted increasing interests for the charge density wave phenomena and possible ferromagnetism. Here, we report on the multiphase behavior and gap opening in monolayer VTe$_2$ grown by molecular beam epitaxy. Scanning tunneling microscopy (STM) and spectroscopy study revealed the (4×4) metallic and gapped (2$\sqrt{3}$×2$\sqrt{3}$) charge-density wave (CDW) phases with an energy gap of ~40 meV. Through the in-plane condensation of vanadium atoms, the typical star-of-David clusters and truncated triangle-shaped clusters are formed in the (4×4) and (2$\sqrt{3}$×2$\sqrt{3}$) phases respectively, resulting in different surface morphologies and electronic structures as confirmed by density functional theory (DFT) calculations with on-site Coulomb repulsion. The CDW-driven reorganization of the atomic structure weakens the ferromagnetic superexchange coupling and strengthens the antiferromagnetic exchange coupling on the contrary, suppressing the long-range magnetic order in monolayer VTe$_2$. The electron correlation is found to be important to explain the gap opening in the (2$\sqrt{3}$×2$\sqrt{3}$) phase.

Two-dimensional (2D) transition-metal dichalcogenides (TMDs) are recently attracting great attention for their diverse physical properties such as ferromagnetism[1,2], Mott-insulator phase[3,4], quantum Hall effect[5-7], superconductivity and charge density wave (CDW)[8-11]. CDW is a periodic symmetry-lowering redistribution of charge in a material, accompanied with periodic modulation of the atomic lattice and electronic densities. The CDW transition in the group VB dichalcogenides, such as $VS_2$, $VSe_2$, $NbSe_2$, $TaS_2$, $TaSe_2$ and $TaTe_2$, has been extensively studied[12-20]. The phase transition exhibits a layer-dependent behavior, for instance, $VSe_2$ and $NbSe_2$ have increased CDW transition temperatures as the thicknesses reduce down to a few monolayers. In monolayer $VSe_2$ and $NbSe_2$, the $T_{CDW}$ can be enhanced to 140 K and 145 K compared to 110 K and 33 K of the bulk $T_{CDW}$ respectively[21-24]. The intricate interplay between CDW and other collective electronic states such as superconductivity has also been explored in $NbSe_2$ and $TaS_2$[25,26]. While the driving mechanisms of CDW in two-dimensional TMDs consisting of Fermi surface nesting[27,28], saddle point singularities[29,30] and electron-phonon coupling[31-34] have been proposed, a consistent picture on the CDW formation is still to emerge.

As for $VTe_2$, CDW phase transition has been observed in $VTe_2$ thin films as the change of resistance at 240 K and 135 K, which can be driven by an electric field due to local Joule heating[35]. When the thickness reduces from multilayer to single layer, the absence of interlayer interaction result in the structural transition from a monoclinic distorted 1T′ structure to a hexagonal 1T structure[36]. The 1T′ structure has a (3×1) double zigzag chain-like modulation of V atoms, which arise from the in-plane metal coordination through d orbital hybridization. As the temperature is cooled below $T_{CDW}$, the monolayer 1T-$VTe_2$ exhibits a (4×4) superstructure, which has been observed by low-energy electron diffraction and scanning tunneling microscopy (STM)[37,38]. By means of angle-resolved photoemission spectroscopy, partial gap-opening with an anisotropic feature was observed in (4×4) reconstructed $VTe_2$ monolayers[39]. Besides CDW phenomena, monolayer $VTe_2$ was also considered as a candidate for realizing 2D ferromagnetic (FM) order[40]. Room-temperature ferromagnetism has been observed in $VTe_2$ nanoplates[2]. However, FM order was absent in monolayer $VTe_2$ according to recent experimental studies[36,37]. The interplay between the long-range charge and spin orders in monolayer VTe$_2$ is yet unknown.

Here we investigate the different CDW phases in monolayer VTe$_2$ by STM/STS and DFT calculations. The metallic (4×4) phase and the insulating (2√3×2√3) phase were revealed to coexist in the different regions of monolayer VTe$_2$ grown on graphite. The microscopic formation mechanism of the (4×4) and (2√3×2√3) superstructures is interpreted by the different types of condensation of V atoms. The driving force of the structural transition from the (4×4) superstructures to the (4×1) and (5×1) stripe configurations is ascribed to the local strain induced by grain boundaries. The distinct electronic structures induced by the atomic reconstructions were calculated by DFT calculations. And the DFT calculations indicate that the ferromagnetic order can be suppressed by the CDW phases.

**Results**

Monolayer 1T-VTe$_2$ is composed of a layer of hexagonally arranged V atoms sandwiched by Te atoms in an octahedral coordination, as sketched in Fig. 1a. The experiments were carried out on the monolayer VTe$_2$ films grown on a highly oriented pyrolytic graphite (HOPG) substrate by molecular beam epitaxy. The large-scale STM images show the morphology of the monolayer VTe$_2$ sample (Fig. 1b and Fig. S1). Two different types of islands，featuring proximately hexagonal shape with straight edges and round shape with curved edges, have been observed on the surface. The surface of the hexagon-shaped islands is corrugated with hexagonally ordered bright spots corresponding to the (4×4) CDW phase. The round islands exhibit relatively smooth surface topography, which corresponds to the (2√3×2√3) CDW phase. The apparent heights of the two types of VTe$_2$ islands are both about 8.7 Å. The hexagonal superstructure of the (4×4) CDW phase has a periodicity of 13.5 Å, as measured by high-resolution STM at 78 K. Fig. 1d and 1e are the high-resolution STM images of the (4×4) and (2√3×2√3) phases, respectively. The bright protrusions in Fig. 1d are assigned to the Te atoms of the top layer, which display a characteristic of bright lines and triangles (also see Fig. S2). Compared with the (4×4) phase, the (2√3×2√3) phase has much less yield and exhibits different surfaces morphology (Fig.1e). The lattice constant of the unit cell marked in Fig. 1e is 11.5 Å. Different from the STM images measured at 78 K, only primitive hexagonal lattice of VTe$_2$ appears without any superstructures on the STM images measured at 300 K. Therefore, the possibility that the superstructures are the moiré pattern between the monolayer VTe$_2$ and HOPG can be ruled out. Nevertheless, when the sample was in-situ heated to 140 K, the CDW structures were still observed in the STM experiments. It implies the CDW transition temperature lying in the range from 140 K to 300 K, consistent with the previously reported value[39].

DFT calculations were carried out to reveal the lattice distortion in the monolayer CDW phases. The reconstructions of V atoms with concomitant displacements of Te atoms emerged after the structural optimizations of the 1T-VTe$_2$ supercells. In the (4×4) supercell, the lattice is deformed into the unit of 13 V atoms with 12 atoms displaced toward the center V atom in the shape of a star of David (Fig. 2a), reminiscent of the (13√3×13√3) superstructure of 1T-TaS$_2$[41]. In the David-star, the neighboring V-V distances are 4-8% shrunk compared to that of the calculated (1×1) unit cell. Consequently, vertical bulge takes place for the topmost Te atoms in the David star, which are 0.16 Å higher than the outside ones according to our calculation. The lattice distortion brings the charge concentration at the center of the David-star, as resolved by the atomic-resolution STM image (Fig. 2b). The feature that three central Te atoms are at the highest intensity with surrounding three Te atoms at less intensity can be reproduced by the simulated STM images (Fig. 2c). As for the (2√3×2√3) supercell, truncated triangle-shaped clusters formed by the condensation of 12 V atoms (Fig. 2d). The average distance between the neighboring vanadium atoms in the truncated triangle-shaped cluster is calculated to be 3.35 Å, which is 4% shortened compared to that of the calculated (1×1) unit cell. The CDW formation energy ΔE has been calculated, which is defined as $E_{nor}-E_{CDW}$, the difference of the total energies between the normal phase and the CDW phase. The CDW formation energies were calculated to be 0.75 eV for the (4×4) phase and 0.34 eV for the (2√3×2√3) phase. The relatively smaller formation energy of the (2√3×2√3) phase implies that the formation of the (2√3×2√3) superstructures is more difficult than the (4×4) phase, which is consistent with the much less yield of the (2√3×2√3) phase on the samples. The atomic-resolution STM image of the (2√3×2√3) phase revealed a flower-like feature that one central Te atom and six surrounding atoms are at the highest brightness (Fig. 2e), which agrees well with the simulated STM image shown in Fig. 2f.

The local strain has been reported to govern the phase transition from the triangular CDW phase to the stripe phase in $NbSe_2$.[42] The grain coalescence can result in the formation of imperfect lattice along the boundaries (Fig. S3). It is believed that in-plane strain exists at the grain boundaries due to the existence of imperfect lattice[43]. Our high-resolution STM image (Fig. 3a) indicates the coexisting of the (4×4) hexagonal superstructure and (4×1) stripe phase at the vicinity of grain boundary. The characteristic of the stripe structure that one atomic chain is brighter than the other three can be reproduced in our simulated STM image, when a moderate 5% lattice strain is applied (Fig. 3b). In addition, (5×1) stripe structure was also observed near domain boundaries, as shown in Fig. 3c and 3d. The spacing between the atomic rows is inhomogeneous which might be inherently associated with the rearrangement of V atoms. The above result implies the crucial role of local strain in the formation of stripe phases.

The local density of states of the monolayer $VTe_2$ was explored by using scanning tunneling spectroscopy (STS). The (4×4) and (2√3×2√3) CDW phases exhibit distinct differences in the dI/dV spectra. As for the (4×4) phase, a non-zero tunneling conductance at the Fermi level was revealed (Fig. 4b), indicating a metallic feature. But in the (2√3×2√3) phase, a CDW gap of ~40 meV can be clearly identified (Fig. 4g). On both sides of the gap, two prominent peaks at -0.13 eV and +0.19 eV are resolved (Fig. 4f). The basic characteristics of the STS curves remain spatially uniform except that the intensity of the peaks varies along with different positions, which can be confirmed by the STS linear mapping (Fig. S4). The different electronic properties of two types of CDW phases can be ascribed to the *p-d* overlap and rehybridization of V and Te orbitals due to the lattice distortion. The strong electron-electron Coulomb interaction localizes the unpaired 3$d$ electrons of V$^{4+}$ ions, leading to the electronic reconstructions. The band structures captured by DFT calculations including the on-site Hubbard U correction also exhibit the metallic and insulating characteristics in the (4×4) phase and (2√3×2√3) phase, respectively (Fig. 4d and Fig. 4h). A quantitatively equal band gap in the (2√3×2√3) phase is reproduced in the calculations at $U$ = 2.0 eV which strongly suggests the crucial role of electron correlation in the formation of insulating electronic state. Both the Fermi surface nesting and electron correlation can contribute to the gap in the (2√3×2√3) phase. The spin-polarized calculations were also employed to detect the magnetic properties of monolayer VTe$_2$, which has been theoretically predicted to exhibit ferromagnetic behavior due to strong electron coupling in the 3$d^1$ odd-electronic configuration of V$^{4+}$. However, the ferromagnetic ordering are both suppressed in the (4×4) and (2√3×2√3) superstructures.

To obtain further insight into the magnetism in the (4×4) and (2√3×2√3) superstructures, we explored the magnetic evolution of undistorted 1T structure under strain. The energy difference between the antiferromagnetic and ferromagnetic states is defined as $\Delta E = (E_{AFM} − E_{FM})$. The calculated ΔE under different biaxial strain is shown in Fig. 5a. The $\Delta E$ is found to get larger as the strain increases. When the strains are set within the range of –7% to –2%, the $\Delta E$ is almost zero. On the other hand, the distance between the nearest-neighboring V atoms $d_{V-V}$ proportionally increases with the increasing strain. While the bond lengths between the V and Te $d_{V-Te}$ increases slowly with increasing strain, as shown in Fig. 5b. Based on the different behaviors of $d_{V-V}$ and $d_{V-Te}$ under strain, the mechanism of the transition between the ferromagnetic and antiferromagnetic states can be explained by the competition between two different exchange interactions[44]. The direct exchange interaction (denoted as $J_D$) cannot be neglected due to the small $d_{V-V}$, which results in the antiferromagnetic state (shown in Fig. 5c). Additionally, the two nearest-neighboring V atoms are connected by a Te atom, thus a superexchange interaction (denoted as $J_S$) is also responsible for the magnetic state (shown in Fig. 5d). Since the V-Te-V bond angle is close to 90°, according to the Goodenough-Kanamori-Anderson (GKA) rules[45,46], the V-Te-V superexchange interaction is ferromagnetic and the $J_S$ is positive. Thus, the magnetic coupling of VTe$_2$ monolayer is depended on the balance of $J_D$ and $J_S$. For the VTe2 monolayer without strain, the $J_S$ is dominated, the system is FM state. When the compressive strain is applied, as shown in Fig. 5b, $d_{V-V}$ and the V-Te-V bond angle decreases, which results in the increasing of $J_D$ and decreasing of $J_S$. At the critical strain of the –2%, the $J_s$ is equal to $|J_D|$, $J_D + J_S=0$. When the $d_{V-V}$ changes from 3.23 Å to 3.41 Å and the V-Te-V bond angle is from 74.5° to 78.974° which corresponds the variation of the strain from –7% to –2%, the $J_D + J_S$ is almost zero, where ferromagentic order of the system disappears. When the compressive strain further increases, the V-Te-V bond angle further decreases, which is much smaller than 90°, the positive $J_S$ turns into the negative value, the antiferromagnetic interaction sharply increases, As shown in Fig. 5a. It would be interesting to study the geometric frustration in this case. As for the (4×4) and (2√3×2√3) superstructures, the formation of the V clusters results in the V-V distance decreasing to about 3.3 Å. The competition between the direct and superexchange interactions will lead to the disappearance of the ferromagnetic order.

**Conclusions**

We successfully fabricated the monolayer VTe$_2$ on the HOPG substrate by the MBE methods. The different CDW phases were directly observed by low-temperature STM. The STM/STS study combined with the DFT calculations unveiled the metallic (4×4) CDW phase and the (2√3×2√3) CDW phase with a gap opening of ~40 meV. The V atoms are displaced into the star-of-David clusters and truncated triangle-shaped clusters in the (4×4) and (2√3×2√3) phases respectively, which are accompanied with the vertical bulge of the Te atoms. The distinct electronic structures induced by the atomic reconstructions were revealed by DFT calculations. And the atomic reconstruction would suppress the ferromagnetic ordering in (4×4) and (2√3×2√3) phases. In addition, the strain induced by grain boundaries can drive the structural transition from the (4×4) superstructure to the (4×1) and (5×1) stripe configurations. Our findings provide a microscopic view to understand the relationship between the atomic reconstruction and electronic properties and imply that the periodic lattice distortion plays a vital role in the formation of CDW phases in TMDs.

**Methods**

**Experimental measurement.** The monolayer $VTe_2$ was synthesized by molecular beam epitaxy in an ultrahigh vacuum (base vacuum $1\times 10^{-10}$ mbar). The HOPG substrate was cleaved and transferred into the UHV chamber to degas at 500 °C The tellurium atoms were sublimed from a quartz crucible at sublimation temperature of 290 °C. Vanadium atoms were evaporated from an electron-beam evaporator. The HOPG substrate was kept at 280 °C during the codeposition of V and Te atoms. Samples was heated by a direct current tungsten filament located on the back side of the sample holder and the sample temperature was measured with a thermocouple. STM measurements were performed on an Omicron low-temperature STM operated at 78 K. An electrochemically etched tungsten tip was used for topographic and spectroscopic measurements. The STM images were taken in the constant-current mode and the voltages refer to the bias on samples with respect to the tip. The dI/dV spectra were acquired by a lock-in amplifier while the sample bias was modulated by a 553 Hz, 10 mV (r.m.s.) sinusoidal signal under open-feedback conditions. The tip state was checked via the appearance of the characteristic Shockley-type surface state on clean Au(111) surfaces.

**Theoretical calculation.** Spin polarized DFT calculations were performed using the periodic plane-wave basis Vienna ab-initio Simulation Package (VASP) code[47,48]. For geometry optimizations and electronic structure calculations, the Perdew-Burke-Ernzerhof (PBE) functional was applied. The valence-core interactions are described by using the Projector Augmented Wave (PAW) method[49]. The plane-wave energy cutoff used for all calculations is 400 eV. The convergence criterion for the forces of structure relaxations is 0.01 eV $Å^{-1}$. A supercell arrangement was used with a 20 Å vacuum layer to avoid spurious interactions between the monolayer $VTe_2$ and periodic images. The electronic structure calculations were performed using a k-point grid of 4 ×4×1. Before the structural optimization of 1T-VTe2 supercell, the small random displacements were performed to several atoms to remove the initial symmetries. The on-site Coulomb repulsion was considered in the DFT calculations.

**Data availability.** The data that support the findings of this study are available from the corresponding author upon reasonable request.

**Acknowledgements**

This work was financially supported by NSFC (Project No. 11374374, 11574403, 21425310) and the computation part of the work was supported by National Supercomputer Center in Guangzhou. C.W. and D.Y. Thank the support by NKRDPC Grants No. 2017YFA0206203, No. 2018YFA0306001, NSFC-11574404, NSFC-11974432, NSFG-2019A1515011337, and Leading Talent Program of Guangdong Special Projects.

**Author contributions**

D.Z. conceived the project; L.M. and L.Z. performed the STM experiments; L.M., W.C. and Y.D. conducted the theoretical calculations. All authors discussed the results and contributed to the final manuscript.

**Additional information**

Supplementary information is available in the online version of the paper. Reprints and permissions information is available online at www.nature.com/reprints. Correspondence and requests for materials should be addressed to D.Y. and D.Z..

**Competing financial interests**

The authors declare no competing financial interests.

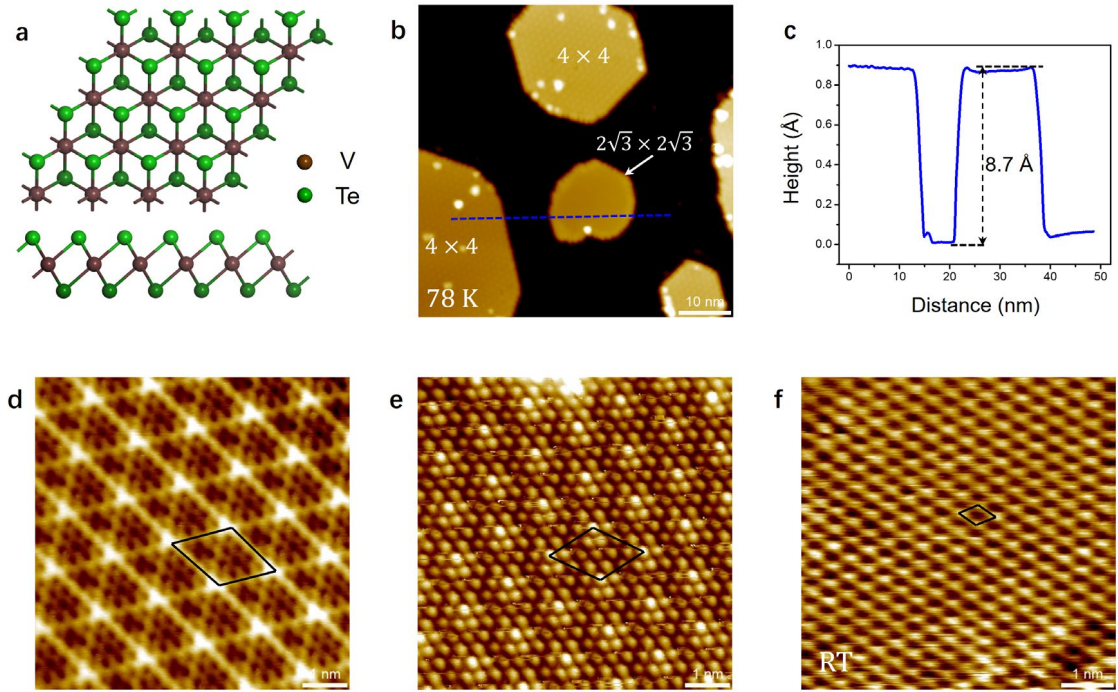

**Figure 1 | Structural model and STM images of monolayer 1T-VTe2 grown on HOPG.** (**a**) Atomic structure of monolayer 1T-VTe2. (**b**) STM image of the monolayer VTe$_2$ measured at 78 K ($V = -2.6$ V, $I = 0.01$ nA). (**c**) The height profile along the blue line in **b**. (**d**) High-resolution STM image of the (4×4) CDW phase ($V = -10$ mV, $I = 2.8$ nA). (**e**) High-resolution STM image of the (2√3×2√3) CDW phase ($V = -0.1$ V, $I = 1.2$ nA). (**f**) High-resolution STM image of monolayer VTe$_2$ measured at room temperature showing a hexagonal lattice ($V = -10$ mV, $I = 3.0$ nA).

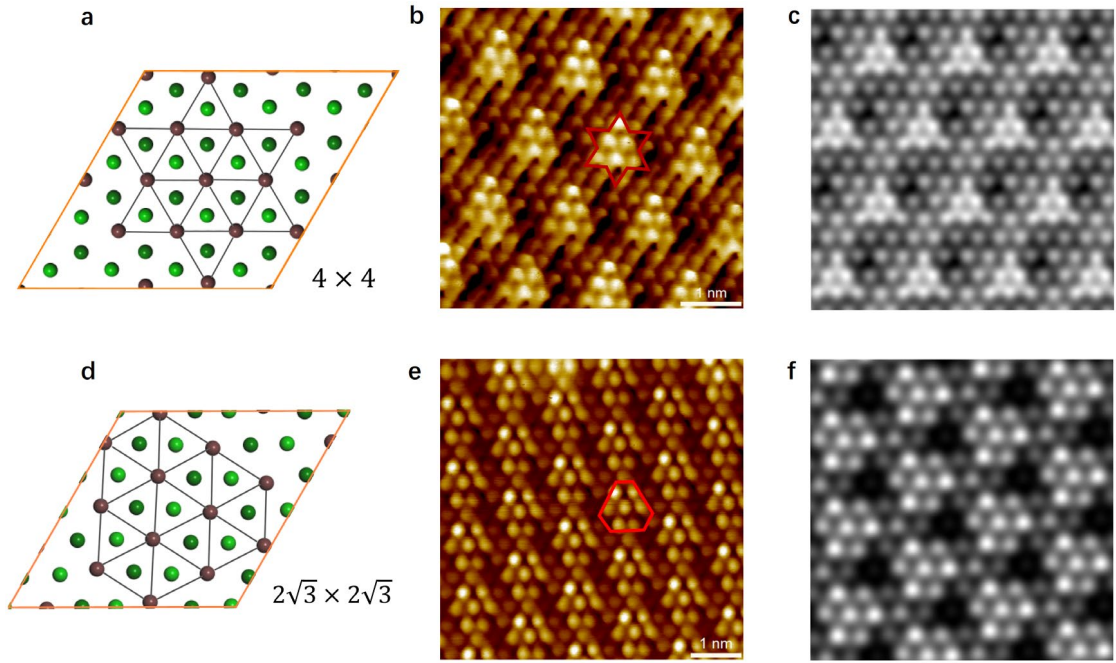

**Figure 2 | Atomic structures of the (4×4) and (2√3×2√3) CDW phases. (a)** Lattice structure of the (4×4) phase exhibiting a star-of-David cluster. **(b)** Atomic resolved topographic image of the (4×4) phase with the David-star pattern (red lines) superimposed ($V = -0.3$ V, $I = 0.8$ nA). Each bright spot on the surface corresponds to a Te atom in the topmost layer. **(c)** Simulated STM images of the (4×4) superstructures. **(d)** Lattice structure of the (2√3×2√3) phase showing a truncated triangle-shaped cluster. **(e)** Atomic resolved STM image of the (2√3×2√3) phase with the overlaid hexagon pattern ($V = -10$ mV, $I = 2.0$ nA). **(f)** Simulated STM image of the (2√3×2√3) superstructures.

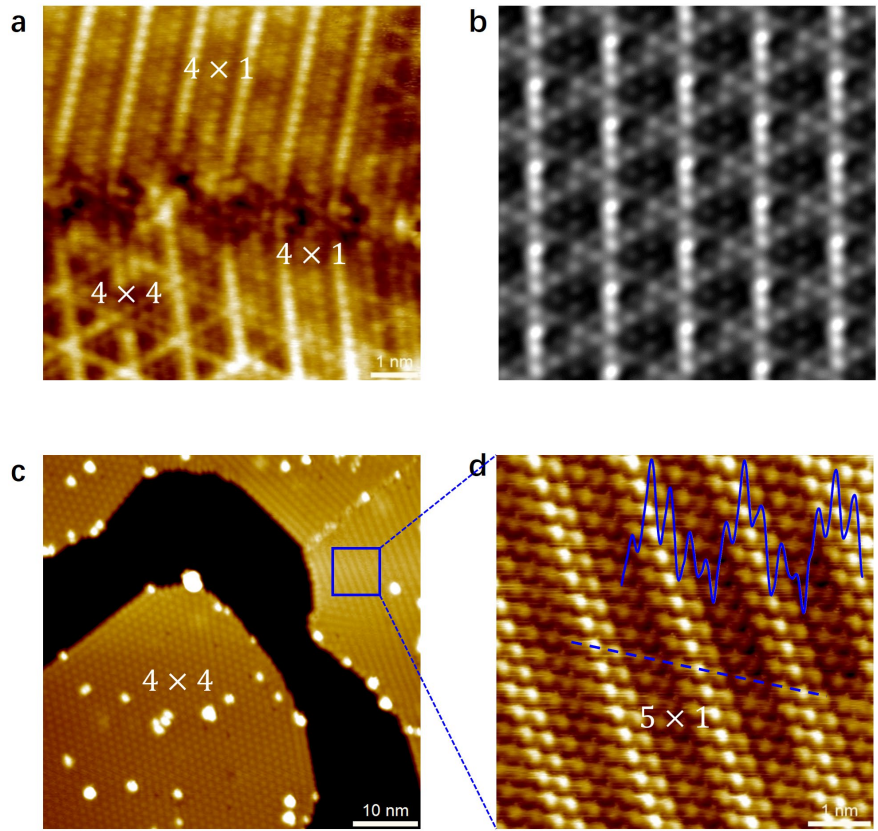

**Figure 3 | Stripe phases induced by the local strain.** (**a**) STM image of the (4×1) stripe phase and (4×4) phase coexisting at the vicinity of grain boundary ($V = -10$ mV, $I = 2.0$ nA). (**b**) Simulated STM image of the stripe structure with 5% lattice strain applied. (**c**) Large-scale STM image of the (5×1) stripe phase located at the grain boundary ($V = -2.4$ V, $I = 0.01$ nA). (**d**) Atomic structure of (5×1) stripe phase ($V = -10$ mV, $I = 2.0$ nA). The line profile corresponding to the blue line reveals a periodicity of 1.7 nm

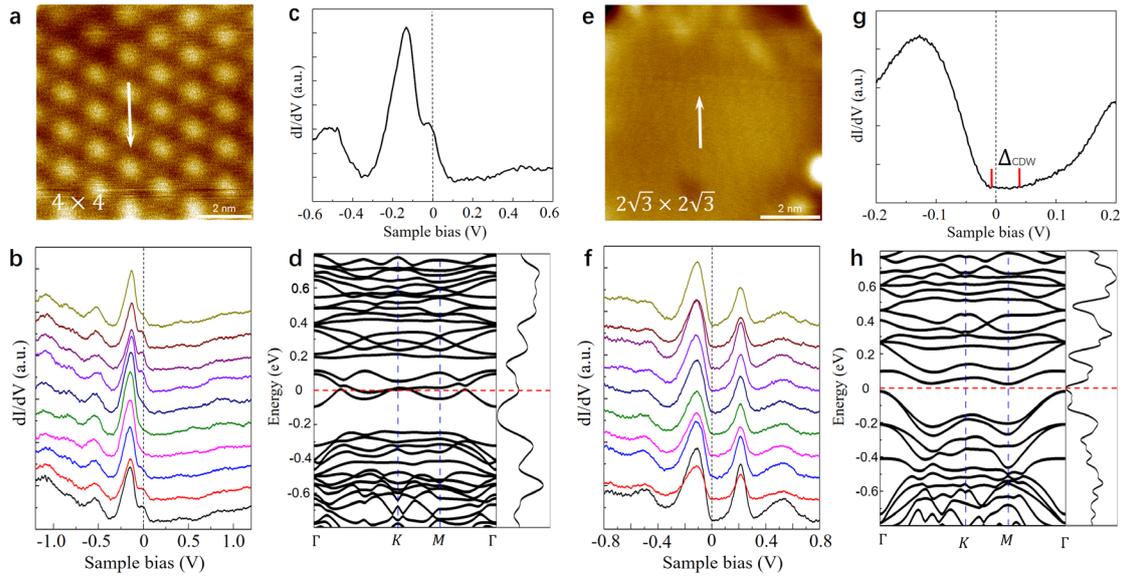

**Figure 4 | Electronic strucutres of the (4×4) and (2√3×2√3) phases.** (**a**,**e**) STM images of the (4 × 4) and (2 √ 3 × 2 √ 3) phases respectively. (**b**,**f**) Differential conductance (dI/dV) spectra taken at different points along the white lines marked in panel a and e respectively. (**c**,**g**) The enlarged dI/dV spectra acquired from the (4×4) and (2 √3×2 √3) phases respectively. A CDW gap of ~40 meV was observed in (2 √3×2 √3) phase. (**d**,**h**) Band structures and corresponding density of states of the (4×4) and (2 √3×2 √3) phases.

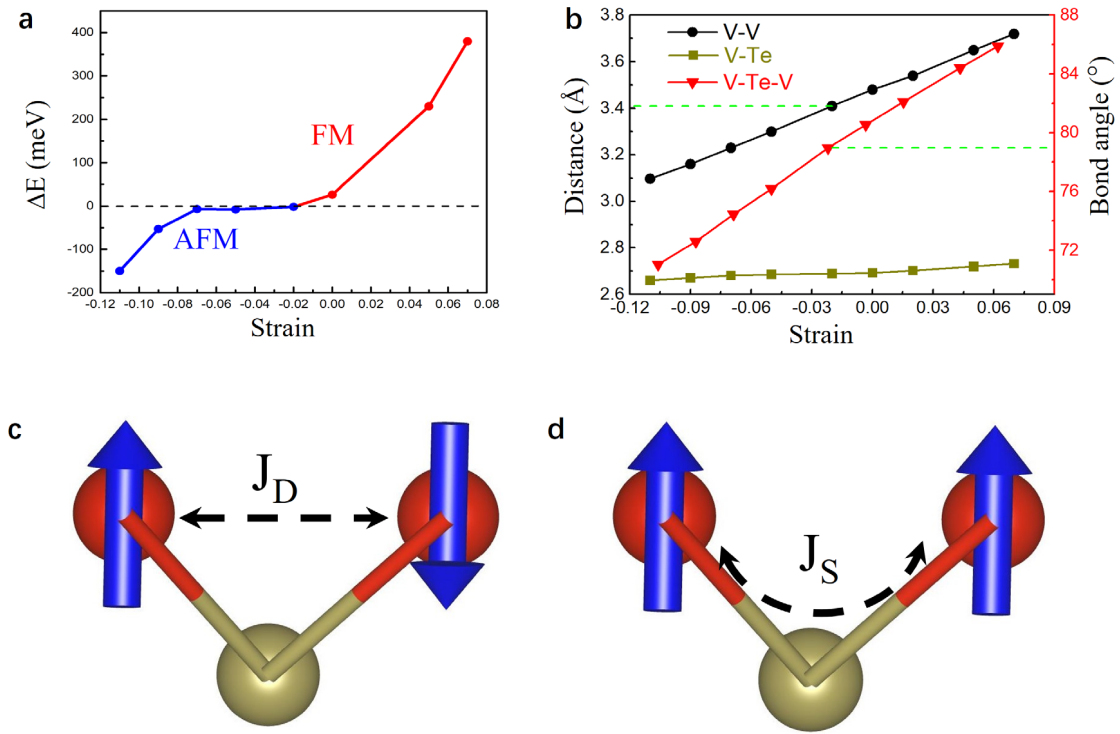

**Figure 5 | Magnetic evolution of undistorted 1T structure under stain**. (**a**) The energy difference between AFM and FM states under different strain. (**b**) The variation of distances of V-V, V-Te and bond angles of V-Te-V as a function of applied strain. (**c**) Schematic showing of the V-V direct exchange interaction and V-Te-V superexchange interaction (**d**) in VTe2 monolayer.